\documentclass[twocolumn,a4paper,floatfix,superscriptaddress]{revtex4}
\usepackage{graphicx}
\usepackage{hyperref}
\usepackage{amsmath}
\usepackage{lineno}
\usepackage{verbatim}
\usepackage{amssymb}
\usepackage[normalem]{ulem}
\usepackage{titlesec}

\newcommand\Ueff{U_{\rm eff}}
\newcommand\UU{U_{m_{1} m_{2} m_{3} m_{4}}}
\newcommand\dd{\rm{d}}
\newcommand\rr{\mathbf{r}}
\newcommand\qq{\mathbf{q}}

\begin{document}

\title{Importance of pressure-dependent electronic interactions and magnetic order on pressure-driven insulator-metal transitions in MnO and NiO}

\author{Bei-Lei Liu}
\affiliation{National Key Laboratory of Computational Physics, Institute of Applied Physics and Computational Mathematics, Beijing 100094, China}

\author{Yue-Chao Wang\footnote{Corresponding authors: yuechao\_wang@126.com}}
\affiliation{National Key Laboratory of Computational Physics, Institute of Applied Physics and Computational Mathematics, Beijing 100094, China}

\author{Yuan-Ji Xu}
\affiliation{Institute for Applied Physics, University of Science and Technology Beijing, Beijing 100083, China}

\author{Xingyu Gao}
\affiliation{National Key Laboratory of Computational Physics, Institute of Applied Physics and Computational Mathematics, Beijing 100094, China}

\author{Hai-Feng Liu}
\affiliation{National Key Laboratory of Computational Physics, Institute of Applied Physics and Computational Mathematics, Beijing 100094, China}

\author{Hai-Feng Song\footnote{Corresponding authors: song\_haifeng@iapcm.ac.cn}}
\affiliation{National Key Laboratory of Computational Physics, Institute of Applied Physics and Computational Mathematics, Beijing 100094, China}

\begin{abstract}
The pressure-driven insulator-metal transition is a crucial topic in condensed matter physics. 
However, even for the prototypical strongly correlated system, NiO, the critical pressure for transition remains debated.
In this work, we evaluated the electronic interactions over a wide range of pressures based on our developed doubly-screened Coulomb correction method and investigated the effects of pressure-dependent electronic interactions and their interplay with magnetic order on the transition. As a validation of the method, we also performed calculations on MnO. The results show that the hybrid functional combined with pressure-dependent screening parameters reasonably describes the insulator-metal transition in MnO. The insulating band gap of antiferromagnetic (AFM) NiO also match well with experiments in both trend and value, which is better than the method using fixed parameters. Further calculations considering magnetic order indicate that as the electronic interactions weaken under pressure, the AFM state of NiO will no longer be stable, a phenomenon that was not observed in previous works. In addition, the results show that, compared with DFT+$U$ within the on-site Coulomb correction framework, the hybrid functional provides a more accurate description of the properties of MnO and NiO at high pressures, highlighting the key role of non-local effects. Our work provides a possible explanation for the long-standing discrepancies in NiO and offers guidance for the development of first-principles methods for correlated electron systems under pressure.

\end{abstract}
\maketitle


\section{Introduction}
\label{sec:Intro}
The pressure-driven insulator-metal transition is a fundamental issue in strongly correlated systems \cite{Mott1937,Boer1937,Mott1990,Imada1998}. The evolution of the electronic structure associated with the insulator-metal transition under pressure is crucial for understanding high-temperature superconductivity \cite{Zhou2022,Ji2024} and heavy fermion behavior \cite{Browne2020,Xu2022}. It also provides key insights for studies of planetary interiors \cite{Greenberg2018} and the design of novel electronic devices \cite{Shao2018}. Various mechanisms for insulator-metal transitions in strongly correlated systems have been proposed. However, the sophisticated interplay between charge, spin, orbital, and lattice degrees of freedom makes understanding the physical mechanisms of specific materials difficult.

Transition metal monoxides such as NiO and MnO, with partially filled $3d$ electrons, are representative materials in strongly correlated systems. While the properties of these materials at ambient pressure have been well understood, discrepancies remain between experimental and theoretical studies on their pressure-driven insulator-metal transitions. For example, experimental studies indicate that NiO undergoes a first-order insulator-metal transition at around 240 GPa \cite{Gavriliuk2012,Gavriliuk2023}. However, nuclear forward scattering experiment shows that the antiferromagnetic state of NiO is
preserved up to 280 GPa \cite{Potapkin2016}, ruling
out the possibility of a magnetic collapse accompanied with Mott transition. 
On the theoretical side, predictions using methods such as DFT+$U$ \cite{Potapkin2016}, hybrid functional \cite{Feng2004}, and DFT+DMFT \cite{Leonov2016,Leonov2020,Gaifutdinov2024} all suggest insulator-metal transition pressures much higher than 240 GPa. For MnO, experimental studies under high pressure are relatively consistent, with insulator-metal transitions observed at around 90–105 GPa\cite{Patterson2004,Yoo2005,Rueff2005}. Although theoretical work using DFT+DMFT has predicted a high-spin state to low-spin state transition as the driving mechanism for the insulator-metal transition\cite{Kunes2008}, DFT+$U$ and hybrid functional can not simulate this metallization \cite{Kasinathan2006,Kasinathan2007}.

These discrepancies highlight the challenges currently faced by theoretical methods under extreme pressure. 
On the one hand, pressure-induced variation in electronic interactions has a significant impact on electronic structure, which can even directly induce the insulator-metal transition \cite{Gavriliuk2008}. However, existing methods still mostly use fixed electronic interactions parameters, ignoring their pressure dependence. On the other hand, recent theoretical work has revealed the significant influence of long-range magnetism on critical pressure and the mechanism of the pressure-driven insulator-metal transition \cite{Gaifutdinov2024}.  
Moreover, the relative stability between different magnetic orders may also be influenced by electronic interactions \cite{Zhao2015}, which have often been overlooked in previous studies.

In this work, we investigate the influence of pressure-dependent electronic interactions and magnetic orderings on the insulator-metal transitions in MnO and NiO using first-principles calculations. To account for the effects of pressure-dependent electronic interactions, we employed our recently developed doubly-screened Coulomb correction (DSCC) method to determine the electronic interaction parameters under pressures. We also examined the enthalpy differences between antiferromagnetic (AFM) and ferromagnetic (FM) orderings to assess the stability of these magnetic states. Our results help to clarify the controversy surrounding the insulator-metal transition in NiO, and highlight the significance of pressure-dependent electron correlations and magnetic orderings in determining the properties of strongly correlated transition metal oxides.


\section{Results and Discussion}
\label{sec:Res}

\subsection{MnO}

\begin{figure*}[tbp]
\centering
\includegraphics[width=0.98\textwidth]{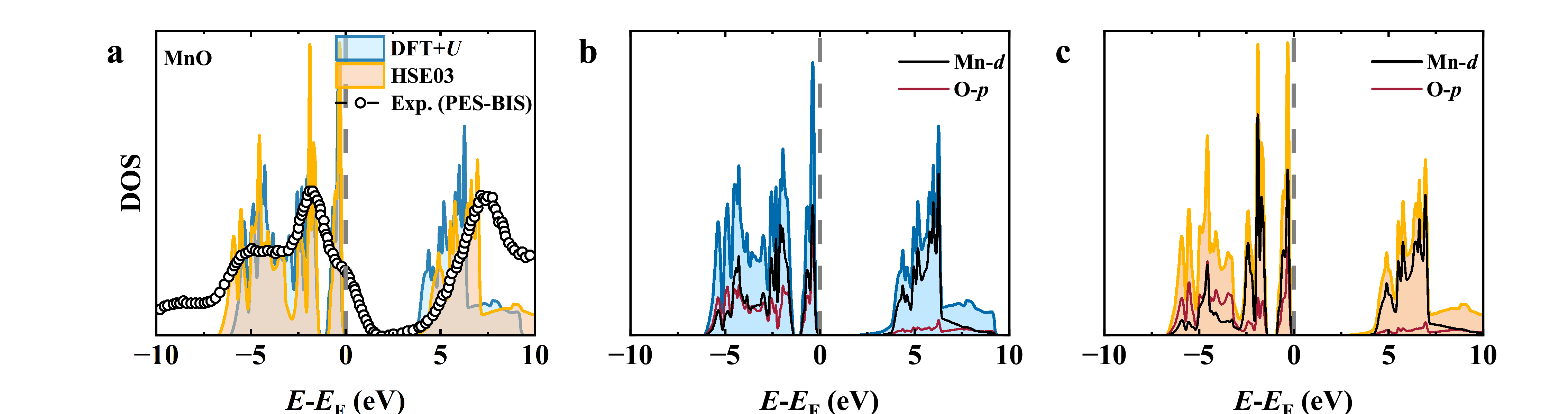}
\caption{(a) Comparison of our theoretical density of states (DOS) of MnO calculated by DFT+$U$ and HSE03 with experimental data. The experimental spectral data measured from the X-ray photoelectron
spectroscopy (XPS) and bremmsstrahlung isochromat spectroscopy
(BIS) are taken from Refs. \cite{Elp1991}. (b) Orbitally-resolved DOS of MnO calculated by DFT+$U$. (c) Orbitally-resolved DOS of MnO calculated by HSE03.}\label{fig:MnO_DOS_P0}
\end{figure*}

MnO exhibits a long-range type-II AFM ordering below the Néel temperature ($T_{N}$=120 K), and a paramagnetic (PM) state at ambient conditions. Our results for the density of states (DOS) of AFM-MnO at ambient pressure by DFT+$U$ with fixed $U,J$ and HSE03 with fixed Hartree-Fock mixing parameter $\alpha$ are shown in Fig.\ref{fig:MnO_DOS_P0} (a)-(c). In agreement with the experiment\cite{Elp1991}, both DFT+$U$ and HSE03 calculations reproduce an insulating state. The highest valence band calculated by DFT+$U$ and HSE03 exhibits O-$p$ features, which is in accord with the charge-transfer insulator nature of MnO \cite{Zaanen1985}. However, it can be observed that the DOS of AFM-MnO near Fermi energy is much sharper than that measured in the PM state at room temperature, which is also consistent with the previous DFT+DMFT study \cite{Mandal2019}.

\begin{figure*}[thbp]
\centering
\includegraphics[width=0.75\textwidth]{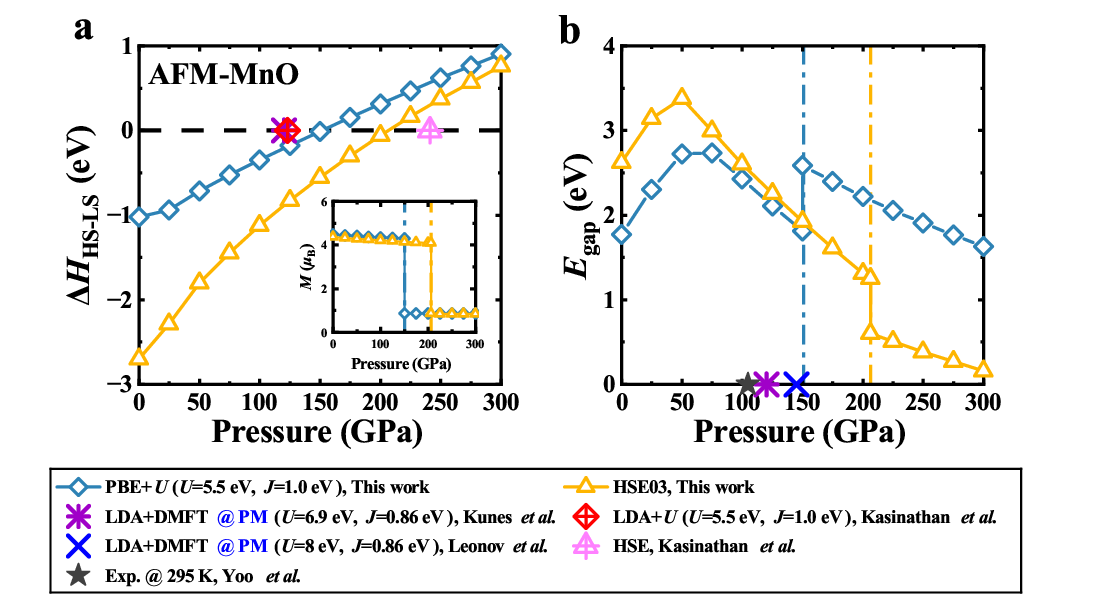}
\caption{(a) Calculated enthalpy difference of high-spin state (HS) respect to low-spin state (LS) of AFM-MnO by DFT+$U$ and HSE03. The critical pressures for HS-LS transition (about 151 GPa of DFT+$U$ and about 206.5 GPa of HSE03) are compared to various theoretical works\cite{Kasinathan2006,Kunes2008}. The inset shows pressure dependence of calculated magnetic moments of AFM-MnO by conventional DFT+$U$ and HSE03.(b) Pressure dependence of calculated band gaps of AFM-MnO by conventional DFT+$U$ and HSE03. The critical pressures for insulator-metal transition are compared to various theoretical works\cite{Leonov2016,Kunes2008} and experimental data\cite{Yoo2005}.}\label{fig:MnO_HSLS} 
\end{figure*}
 
The insulator-metal transition in MnO is observed experimentally at around 90–105 GPa\cite{Patterson2004,Rueff2005,Yoo2005,Kolorenc2007}.
It is proposed that the driving force of the transition is the increase of crystal-field splitting under pressure \cite{Kunes2008}, which causes the redistribution of electrons from the insulating high-spin (HS, $S=\tfrac{5}{2}$) state to the metallic low-spin (LS, $S=\tfrac{1}{2}$) state. Our results for the enthalpy difference between the HS and LS states of AFM-MnO as a function of compression are shown in Fig.\ref{fig:MnO_HSLS}(a). The HS-LS transition pressures predicted by DFT+$U$ and HSE03 are 151 GPa and 206.5 GPa, respectively. The tendency for the transition pressure predicted by DFT+$U$ to be lower than that predicted by hybrid functional is in agreement with previous calculations \cite{Kasinathan2006}. However, the transition pressures obtained from DFT+$U$ and HSE still show discrepancies with previous work. This is probably due to differences in functionals and basis sets. The variation in the magnetic moment under pressure, as shown in the inset of Fig.\ref{fig:MnO_HSLS}, also reflects this transition.

The band gap of AFM-MnO obtained by various approaches as a function of pressure is summarized in Fig.\ref{fig:MnO_HSLS}(b). Similar to the results obtained by previous studies \cite{Kasinathan2006,Kasinathan2007}, both DFT+$U$ and HSE03 capture the HS-LS transition under pressure. However, these approaches produce only an insulating HS-LS transition. This insulating transition is not only inconsistent with the insulator-metal transitions at 0 K obtained by extrapolation based on room-temperature experiment \cite{Yoo2005}, but also differs from the results obtained by DFT+DMFT, which yield insulator-metal transitions within the range of 100-150 GPa \cite{Kunes2008,Leonov2016,Leonov2020}. These DFT+DMFT simulations even use a larger value of $\Ueff$, which would typically lead to larger band gaps. Nevertheless, it should be noted that DFT+DMFT simulations were performed in the PM state of MnO at $T$=1160 K, while the AFM ordering leads to reduced symmetry and weakened hybridization between O-$p$ and TM-$d$ orbitals \cite{Li2023}, which could increase the band gap. 

To investigate the influence of the magnetic order on the changes in the electronic structure under pressure, the band gap and the magnetic moment as a function of pressures of FM-MnO are presented in Fig.S1(a),(b). The DFT+$U$ calculations indicate that FM-MnO is in an insulating state at ambient pressure. The kinetic energy of the system increases under pressure, and Mno transforms into a metallic state at around 50 GPa. Notably, after the HS-LS transition at around 133 GPa, MnO switches back from a metallic state to an insulating state. Kuneš $et~al.$ \cite{Kunes2008} drew an analogy between the insulator-metal transition in MnO and the phase transition in a two-band model under increasing crystal-field splitting \cite{Werner2007}. However, the orbitally polarized insulator phase, which the two-band model predicts at larger crystal-field splitting, is not observed. DOS from DFT+$U$ calculations shown in Figure S2 within the supplementary material exhibits the characteristics of an orbitally polarized insulator. In the AFM phase, the state after the HS-LS transition is also an orbitally polarized insulator. However, due to the influence of magnetic order, the first stage of the kinetic energy-induced insulator-metal transition does not occur. Moreover, in the DFT+$U$ simulation, the band gap of the LS state is contributed by both the on-site electronic interaction and the crystal-field splitting. Therefore, whether in the FM or AFM order, the band gap of the LS state is higher than that of the HS state at the same pressure. This phenomenon does not exist in the hybrid functional calculations with global corrections, and the HS-LS transition of the obtained FM-MnO occurs within the metallic state.

The comparison of simulated results under AFM and FM state shows that although magnetism has little effect on the pressure for the HS-LS transition, it has a significant impact on the simulated band gap. Both DFT+$U$ and HSE03 reproduced the insulator-metal phase transitions under the hypothetical FM state, but these transitions are driven by the bandwidth, which is significantly different from the mechanism revealed by the previous DFT+DMFT simulation \cite{Kunes2008}. In particular, under the FM state, DFT+$U$ also simulated the HS-LS transition accompanied by a metal-insulator transition, in contrast to the transition from an insulating phase to a metal.

\begin{figure}[thbp]
\centering
\includegraphics[width=0.49\textwidth]{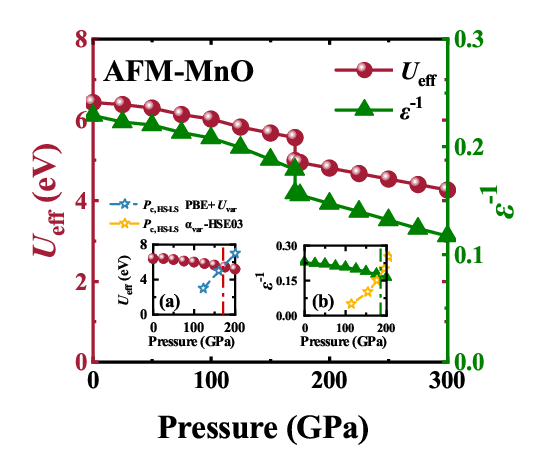}
\caption{Pressure dependence of the calculated effective on-site Coulomb interaction $\Ueff$, and the inverse of static dielectric screening strength $\varepsilon_\infty^{-1}$ of AFM-MnO by DSCC approach. 
The inset (a) shows calculated critical pressures $P_c$ for HS-LS transition using DFT+$U$ with given $\Ueff$. The intersection of the $\Ueff$-$P_c$ curve and the $P$-$\Ueff$ curve is estimated as the critical pressure for the HS-LS transition of DSCC approach (about 171 GPa). The inset (b) shows calculated critical pressures for HS-LS transition using HSE03 with given $\varepsilon$. The intersection of the $\varepsilon$-$P_c$ curve and the $P$-$\varepsilon$ curve is estimated as the critical pressure for the HS-LS transition of $\alpha_{\rm sc}$-HSE approach (about 185 GPa). }\label{fig:MnO_AFM_DSCC}
\end{figure}

\begin{figure*}[thbp]
\centering
\includegraphics[width=0.98\textwidth]{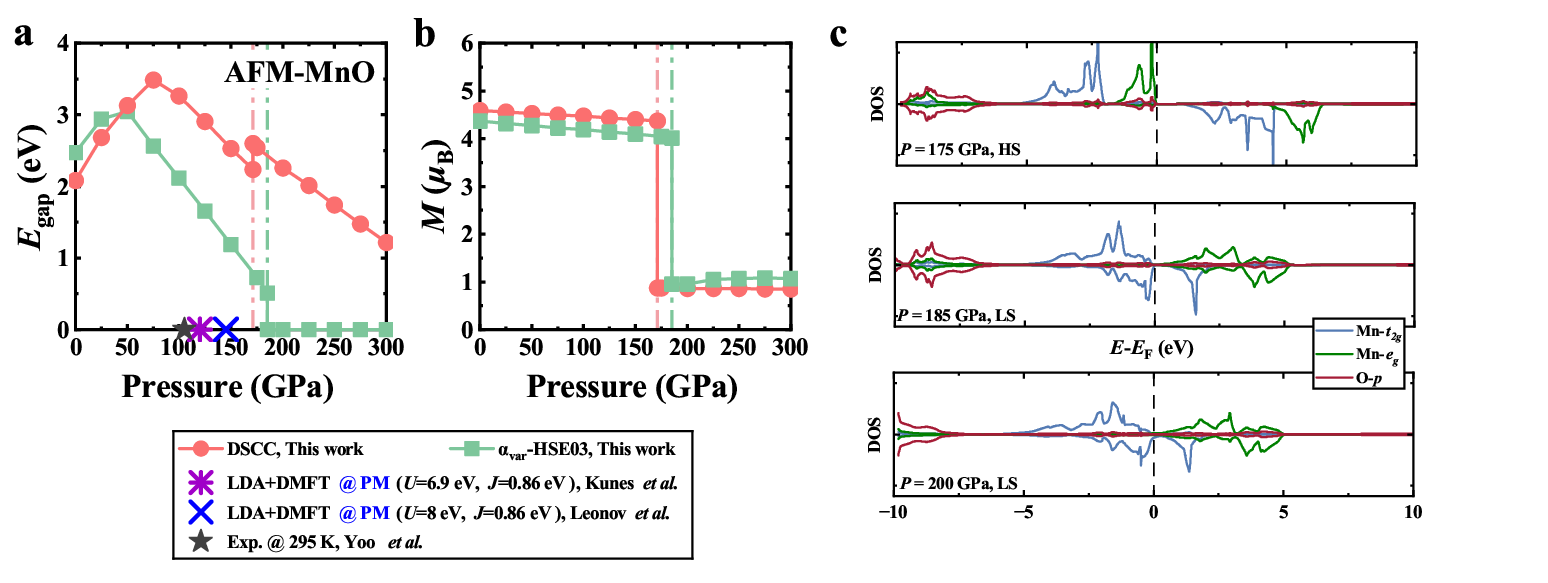}
\caption{ 
Pressure dependence of (a) calculated band gaps and (b) magnetic moments of AFM-MnO by DSCC and $\alpha$-HSE03 approach considering pressure dependent electronic interactions. The critical pressures for insulator-metal transition are compared to various theoretical works\cite{Leonov2016,Kunes2008} and experimental data\cite{Yoo2005}. 
(c) Pressure evolution of orbital-resolved density of states from 175 GPa to 200 GPa, obtained by $\alpha_{\rm sc}$-HSE03 calculations.
}\label{fig:MnO_Egap_Mag}
\end{figure*}

Coming back to AFM-MnO,  
DSCC approach is employed to evaluate the pressure-depedent on-site electronic interaction parameter $U,J$, as well as the inverse of static dielectric screening parameter $\varepsilon_\infty^{-1}$. The results of $\varepsilon_\infty^{-1}$ and the effective on-site electronic interaction strength $\Ueff \equiv U-J$ are shown in Fig.\ref{fig:MnO_AFM_DSCC}. These parameters reflect the screened interactions between localized electrons, which gradually weaken within the range of 0-300 GPa. The estimated critical pressures for HS-LS transition of DFT+DSCC and $\alpha_{\rm sc}$-HSE03-HSE are also depicted in the inset of Fig. \ref{fig:MnO_AFM_DSCC} (a),(b). Differences in electronic interactions caused by changes in electronic structure, such as those associated with the HS-LS transition, can be identified by the DSCC method. There is a difference of approximately 0.5 eV between $\Ueff$ for HS and LS states. In particular, the electronic interaction strength in the LS state is weaker, which is consistent with the previous studies on BiFeO$_3$ and FeBO$_3$ \cite{Gavriliuk2008}. 

The band gaps and magnetic moments calculated by DFT+DSCC considering the pressure-dependent on-site Coulomb interactions $U,J$, and $\alpha_{\rm sc}$-HSE03 considering the pressure-dependent mixing parameter $\alpha_{\rm sc}=\varepsilon_\infty^{-1}$ are shown in Fig.\ref{fig:MnO_Egap_Mag}(a),(b). At pressures above 225 GPa, the band gap simulated by DFT+DSCC is smaller than that obtained by parameter-fixed DFT+$U$. However, the metallic state is still not reproduced within 300 GPa. DFT+DSCC approach is based on the framework of on-site corrections, which can effectively correct Coulomb interactions between electrons within the same site. It lacks corrections for non-local correlations. The hybrid functional method is expected to alleviate this problem by incorporating a certain amount of Fock term, which account for the static exchange interactions between all electron states.  
The $\alpha_{\rm sc}$-HSE03 approach, using the pressure-dependent mixing parameter derived from DSCC, simulated the metallization associated with the HS-LS transition, highlighting the crucial role of pressure-dependent electronic interactions and non-local effect. Fig.\ref{fig:MnO_Egap_Mag}(c) shows the density of states changes in MnO under pressure. At 185 GPa, the band gap vanishes to 0, and the HS-LS transition is clearly observed. The main contributions near the Fermi surface change from $e_{\rm g}$ to $t_{\rm 2g}$ orbital. 

\begin{figure}[thbp]
\centering
\includegraphics[width=0.49\textwidth]{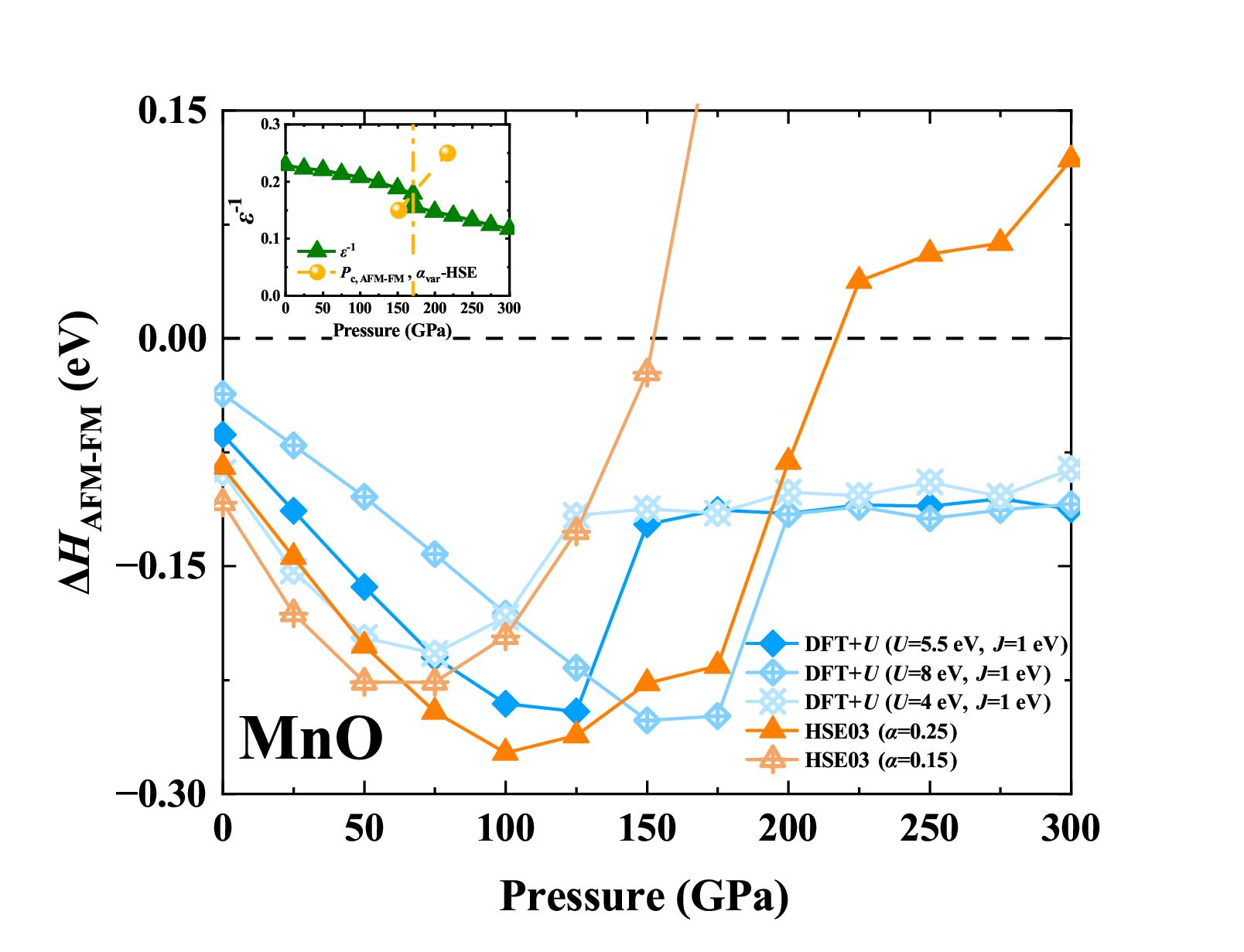}
\caption{Calculated enthalpy difference of AFM state respect to FM state of MnO by DFT+$U$ with different on-site interaction parameters $U,J$ and HSE03 with different mixing parameters $\alpha$. The inset shows calculated critical pressures for AFM-FM transition using HSE03 with given $\alpha$. The intersection of the $\varepsilon$-$P_c$ curve and the $P$-$\varepsilon$ curve is estimated as the critical pressure for the AFM-FM transition of $\alpha_{\rm sc}$-HSE approach (about 171 GPa).  }\label{fig:MnO_AFM_FM}
\end{figure}

Given that the insulating properties between different magnetic states differ under high pressure, we calculated the stability of the AFM state. The enthalpy difference of AFM state respect to FM state of MnO under pressure is presented in Fig.\ref{fig:MnO_AFM_FM}. The AFM state becomes unstable under high pressure, predicted by HSE03, which is consistent with the conclusions of previous work \cite{Archer2011}.  
To investigate the impact of electronic interactions on stability of magnetic states, we performed calculations using different parameters $U,J$, and $\alpha$, but kept the same across the range of pressure. As the mixing parameter $\alpha$ decreases, the critical pressure for the AFM-FM transition gradually decreases. The estimated critical pressure for AFM-FM transition in MnO for $\alpha_{\rm sc}$-HSE03 is presented in the inset of Fig.\ref{fig:MnO_AFM_FM}.
The AFM state is no longer stable at least around 171 GPa. At room temperature, MnO first undergoes an AFM-PM transition, followed by an insulator-metal transition in the PM state \cite{Yoo2005}. The results suggest that at 0 K, the AFM-PM transition also accompanied by an insulator-metal transition (about 180 GPa predicted by $\alpha_{\rm sc}$-HSE03), which is different from the previous LSDA +$U$ simulation that the AFM state is stable up to a high pressure\cite{Kasinathan2007}.

\subsection{NiO}


\begin{figure*}[thbp]
\centering
\includegraphics[width=0.98\textwidth]{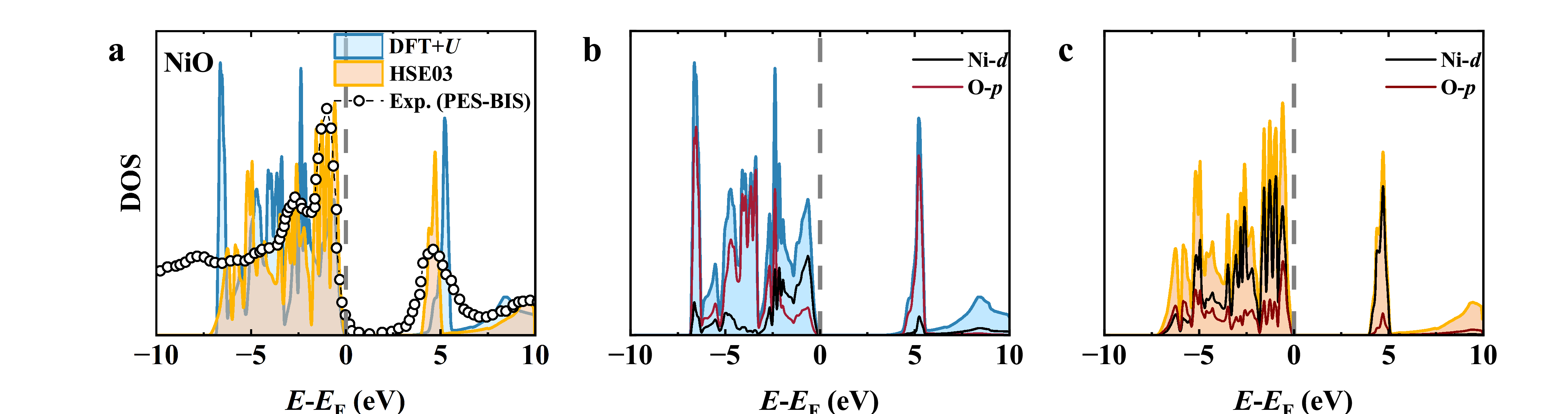}
\caption{(a) Comparison of our theoretical DOS of NiO calculated by DFT+$U$ and HSE03 with experimental data. The experimental spectral data measured from the XPS and BIS are taken from Refs. \cite{Sawatzky1984}. (b) Orbitally-resolved DOS of NiO calculated by DFT+$U$. (c) Orbitally-resolved DOS of NiO calculated by HSE03.}\label{fig:NiO_DOS}
\end{figure*}

In contrast to MnO, the critical pressure for insulator-to-metal transition of NiO remains highly controversial, since significant discrepancies between experimental studies \cite{Gavriliuk2012,Gavriliuk2023,Potapkin2016}, and between experimental and theoretical works\cite{Feng2004,Leonov2016,Leonov2020,Gaifutdinov2024}. We employ approaches considering pressure-dependent electronic interactions, and examine the stability of AFM state.

\begin{figure}[thbp]
\centering
\includegraphics[width=0.49\textwidth]{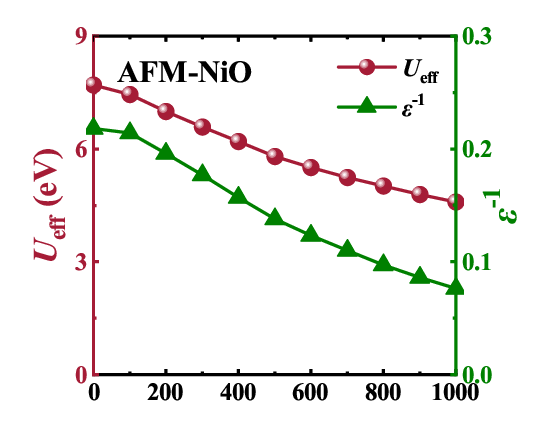}
\caption{Pressure dependence of the calculated effective on-site Coulomb interaction $\Ueff$, and the inverse of static dielectric screening strength $\varepsilon_\infty^{-1}$ of AFM-NiO by DSCC approach.}\label{fig:AFM_NiO_DSCC}
\end{figure}

\begin{figure*}[thbp]
\centering
\includegraphics[width=0.98\textwidth]{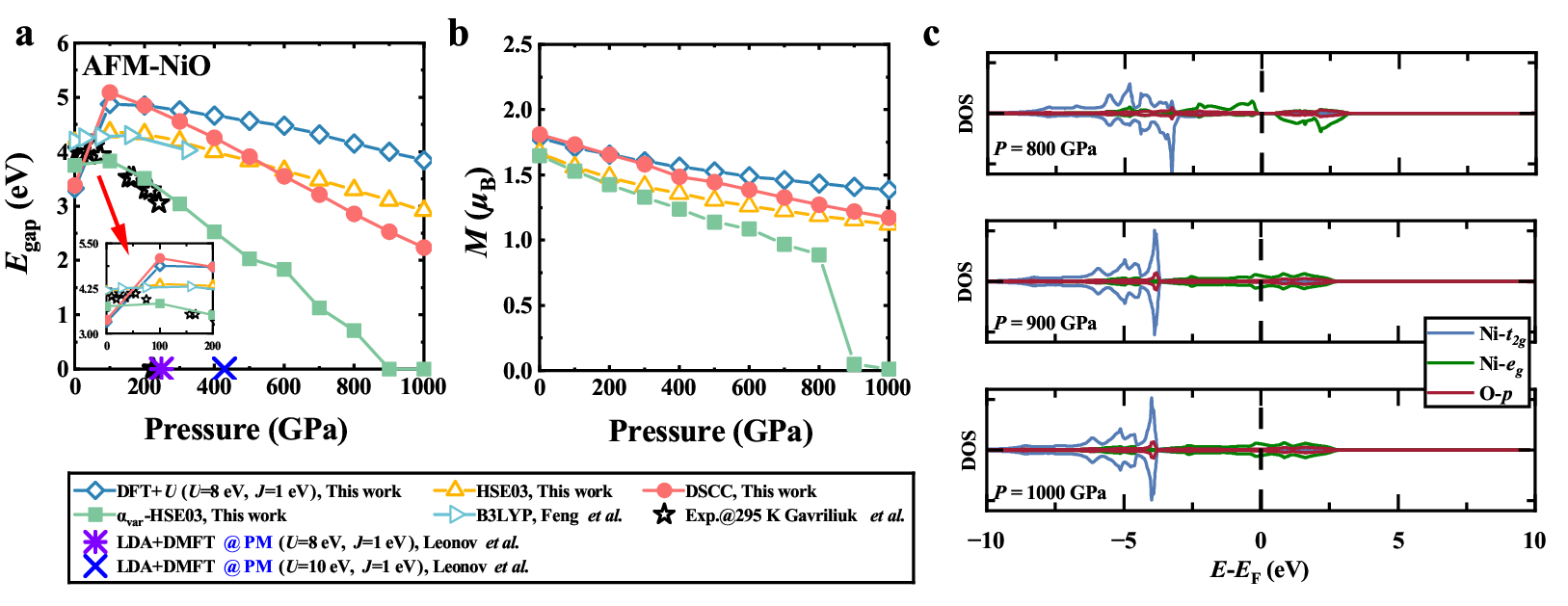}
\caption{ 
Pressure dependence of (a) calculated band gaps and (b) magnetic moments of AFM-NiO by conventional DFT+$U$, HSE03, and DSCC, $\alpha$-HSE03 approach. The band gaps calculated by Feng $et~al.$ \cite{Feng2004} using B3LYP hybrid functional, and measured by Gavriliuk $et~al.$ \cite{Gavriliuk2008} also presented for comparison. (c) Pressure evolution of orbital-resolved density of states from 800 GPa to 1000 GPa, obtained by $\alpha_{\rm sc}$-HSE03 calculations.}\label{fig:AFM_NiO_DSCC_Egap}
\end{figure*}

Our results for the DOS of NiO by DFT+$U$ and HSE03 at ambient pressure are shown in Fig.\ref{fig:NiO_DOS}(a)-(c). DFT+$U$ and HSE03 yield an insulating state in qualitative agreement with the experiments\cite{Sawatzky1984}. HSE03 shows better agreement in peak positions compared to experimental data. Both DFT+$U$ and HSE03 successfully reproduce the charge-transfer insulator nature with O-$p$ characteristics in the highest valence band states, while DFT+$U$ overestimates the O-$p$ weight in these states compared to experiments.

The results for the inverse of static dielectric screening strength $\varepsilon_\infty^{-1}$ and effective on-site electronic interaction $\Ueff$ for NiO evaluated by DSCC approach are presented in Fig.\ref{fig:AFM_NiO_DSCC}(a). The simulated $\varepsilon_\infty^{-1}$ and $\Ueff$ continuously decrease with increasing pressure, without abrupt changes. Our results for the band gap and magnetic moment of AFM-NiO are shown in Fig.\ref{fig:AFM_NiO_DSCC_Egap}(a),(b). The band gap obtained by conventional DFT+$U$ and HSE03 increases from 0 to 100 GPa, and then continues to decrease at higher pressures, while the increase trend from 0 to 100 GPa is more pronounced for DFT+$U$. Band gaps simulated by HSE03 are in agreement with those obtained by previous work using hybrid functional B3LYP \cite{Feng2004}. Compared to experimental measurements \cite{Gavriliuk2008}, band gaps obtained by DFT+$U$ and HSE03 are significantly overestimated above 100 GPa. 

DFT+DSCC approach considering pressure-dependent electron interactions, captures the trend of band gap above 100 GPa that is similar to experiments. However, there is a significant increase from 0 to 100 GPa, and systematic deviations in the band gap in the whole range. The trend and magnitude of $\alpha_{\rm sc}$-HSE approach is agree well with experimental measurement of band gap below 240 GPa. Based on this result, the pressure-dependent electronic interactions are important for capturing the trend of the trend of band gap under pressure in NiO, while the non-local effects are mainly reflected in the prediction of the absolute value of the band gap. However, considering only the AFM state, $\alpha_{\rm sc}$-HSE predicts an insulator-metal transition pressure of around 900 GPa, which is significantly higher than the experimentally reported 240 GPa \cite{Gavriliuk2012,Gavriliuk2023}. Fig. \ref{fig:AFM_NiO_DSCC_Egap} (c) shows the changes in the density of states of AFM-NiO under pressure. Above 900 GPa, the band gap decreases to 0, and the magnetic moment collapses rapidly.
The experimental observation of metallization around 240 GPa may be closely related to a pressure-induced AFM-PM transition, as seen in MnO, which is not fully captured due to the static treatment of magnetism.

\begin{figure}[thbp]
\centering
\includegraphics[width=0.49\textwidth]{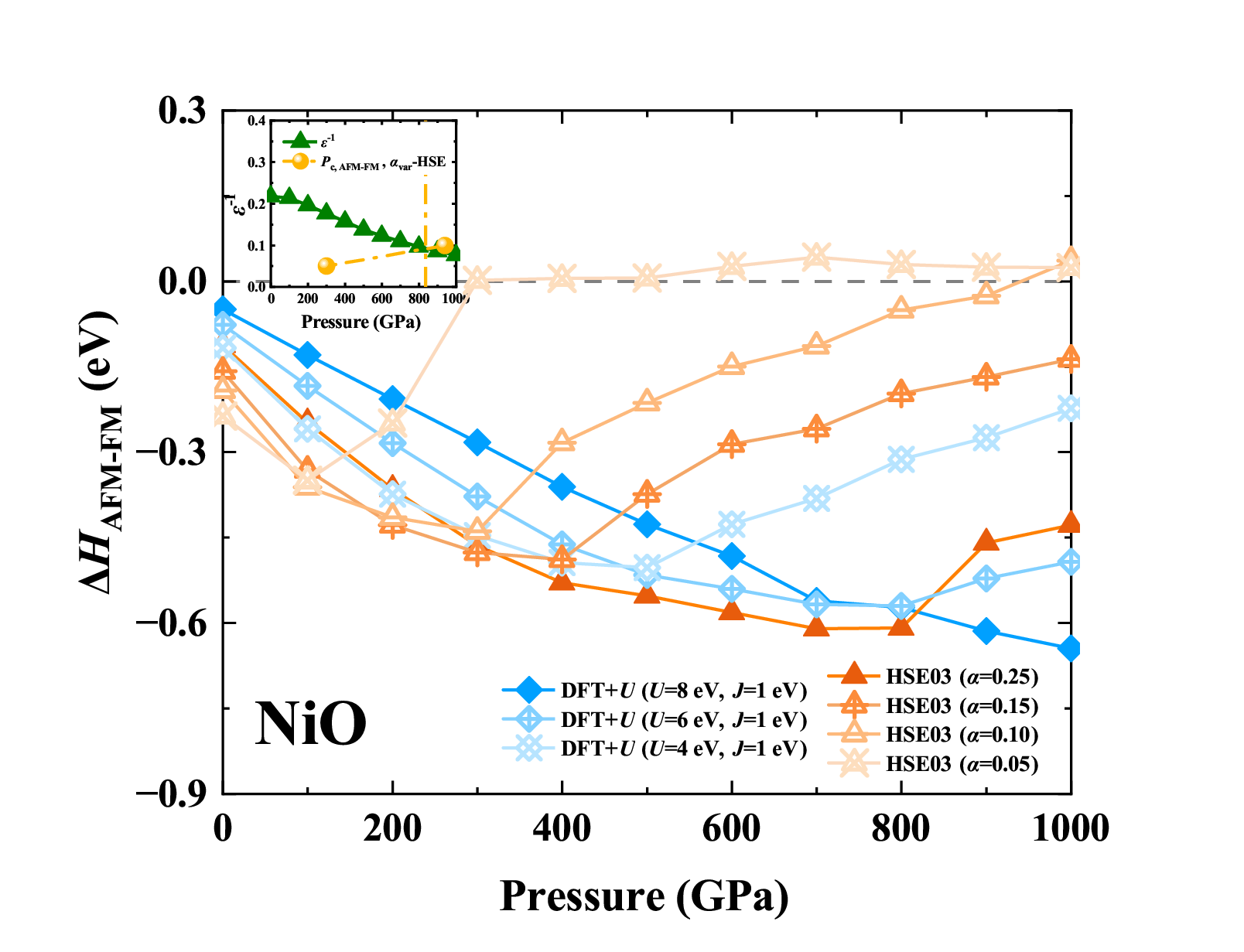}
\caption{Calculated enthalpy difference of AFM state respect to FM state of NiO by DFT+$U$ with different on-site interaction parameters $U,J$ and HSE03 with different mixing parameters $\alpha$. The inset shows calculated critical pressures for AFM-FM transition using HSE03 with given $\alpha$. The intersection of the $\varepsilon$-$P_c$ curve and the $P$-$\varepsilon$ curve is estimated as the critical pressure for the AFM-FM transition of $\alpha_{\rm sc}$-HSE approach (about 836 GPa).}\label{fig:NiO_AFM_FM}
\end{figure}

Further, we calculated the enthalpy differences of the AFM state with respect to the FM state, which are shown in Fig.\ref{fig:NiO_AFM_FM}. Except for DFT+$U$ with $U$=8 eV, the enthalpy differences of the AFM state with respect to the FM state obtained from all simulations show the trend that first decreasing and then increasing with pressure in the range of 0–1000 GPa. In particular, as $U$ decreases and $\alpha$ decreases, the critical pressure at which the enthalpy difference begins to increase gradually decreases. For HSE03 with $\alpha$=0.05, an AFM-FM transition is predicted at $\sim$300 GPa. The estimated critical pressure for AFM-FM
transition in NiO for $\alpha_{\rm sc}$-HSE03 is presented in the inset of Fig \ref{fig:NiO_AFM_FM}. Our results indicate that, as electronic interactions weaken with increasing pressure, the AFM state will no longer be stable at least 836 GPa. Before this, the AFM-PM transition may have occurred. Considering that the insulator-metal transition occurs earlier \cite{Leonov2020} in the PM state, the change in magnetic order is highly critical and may even directly lead to metallization. The pressure dependence of the band gap and magnetic moment simulated for the hypothetical FM-NiO is also shown in Fig.S3 (a),(b) within supplemental material.

On the magnetic order of NiO under high pressure, some explorations have been carried out. Potapkin $et~al.$ \cite{Potapkin2016} showed that NiO still maintains AFM order up to 280 GPa, but in conflict with the observation of the insulator-metal transition around 240 GPa. In terms of theoretical calculations, Archer $et~al.$ \cite{Archer2011} showed that both LDA and PBE calculations have demonstrated a trend of decreasing Néel temperature with increasing pressure. Recent DFT+DMFT simulations at 290 K with fixed $U$ and $J$ parameters predict that the AFM order remains stable and maintains its insulating character up to very high pressures (a compression ratio of 0.4, above 1000 GPa) \cite{Gaifutdinov2024}. As pressure increases, NiO transforms from a Mott insulator into a correlated-assisted Slater insulator (a band insulator driven by long-range magnetic order). However, it was also found that the Néel temperature decreases with lower electronic interaction at a given pressure. To further verify the magnetic state of NiO, we suggest that detailed measurements of magnetic order under pressure could be performed in the future.


\section{Summary}
\label{sec:Conclu}

We investigated the evolution of electronic structure under pressure for the prototypical correlated electron systems MnO and NiO, by DFT+$U$ and hybrid functionals that account for pressure-dependent electron interactions. The results contribute to understanding the long-standing discrepancies among studies of pressure-driven insulator-metal transition for NiO.

First-principles calculations that consider pressure-dependent electronic interactions have yielded an insulator-metal transition in MnO that is close to the results extrapolated from room-temperature experiments, and the insulating band gap obtained for NiO is in good agreement with the trends and values of previous experiment. Our results demonstrate the important role of pressure-dependent electronic interactions in simulating the electronic structure and magnetic states of strongly correlated compounds over a wide range of pressures. We also examined the effects of magnetic orders. The antiferromagnetic phase becomes unstable as electron interactions weaken with increasing pressure. The change in magnetic order occurs prior to the insulator-metal transition and may even drive the transition. In addition, for transition metal oxide systems, hybrid functional calculations provide more accurate descriptions of the properties under pressure compared to DFT+$U$ in the on-site Coulomb correction framework. This highlights the crucial impact of non-local effects.

Our work provides guidance for the development of electronic structure calculation methods under high pressure and offers a possible explanation for the much lower insulator-metal transition pressure observed experimentally in NiO. Since our study has not yet considered the effects of structural distortions and temperature on magnetic order and electronic structure, the obtained transition pressure for NiO is still higher than the experimental value. Future research can further take these factors into account and conduct simulations on the total energy and electronic structure of the paramagnetic state, to clarify the specific mechanism of the insulator-metal transition in NiO.

\section{Method and Computational Details}
\label{sec:Meth}

We employed approaches based on on-site Coulomb correction framework 
and non-local hybrid functional 
in simulations. In the on-site Coulomb correction framework, the ground-state energy can be written as follows: 
\begin{equation}
\label{eq:tot}
E_{\rm tot} = E_{\rm DFT} + E_{\rm Hub} + E_{\rm dc},
\end{equation}
where $E_{\rm DFT}$ refer to the energy of system described by DFT, the double counting term $E_{\rm dc}$ is used to cancel the electron-electron interaction that have already been included in DFT. $E_{\rm Hub}$ described the interaction energy between $d$-electrons form Hubbard model, it takes the following form:
\begin{equation}
\begin{split}
\label{eq:on-site}
&E_{\rm Hub} =\frac{1}{2}\sum_{{m},\sigma}\big[\UU n_{m_1,m_2}^\sigma n_{m_3,m_4}^{\bar{\sigma}}  \\
& +(\UU-U_{m_{1} m_{3}m_{4}m_{2}})n_{m_1,m_2}^\sigma n_{m_3,m_4}^\sigma\big],
\end{split}
\end{equation}
where $\left\{m\right\}$ is the index set of local orbitals, $m_1,m_2,m_3,m_4\in\{m\}$, $\sigma$ stands for the spin, $n_{i,j}^\sigma$ is the occupation matrix element, which are determined by the projection of occupied Kohn-Sham orbitals onto local orbitals \cite{Shick1999,Bengone2000,Amadon2008}. The interaction matrix elements reflect the strength of the screened interactions between the localized electrons on the same shell and the same atomic site, determined by the localization of the electrons and the local screening strength, play a significant role in simulations. However, in conventional DFT+$U$ with fixed $U,J$ parameters, these elements are determined by $U,J$ parameters. These parameters are generally obtained by fitting to experimental data at ambient pressure, and kept as a fixed value within the entire range of pressures. To investigate pressure-dependent electron interactions, we utilized DSCC approach to evaluated interaction matrix elements for $3d$-electrons. In DSCC approach, the doubly screened model dielectric function $\varepsilon(\qq)$ is used to determines the screened Coulomb potential $v_{\rm sc}$:
\begin{equation}
\label{eq:eps}
\varepsilon(\qq) = 1+ \left\{(\varepsilon_\infty - 1)^{-1}+\left(\frac{q}{\lambda_{\rm TF}}\right)^2\right\}^{-1},
\end{equation}
where $\lambda_{\rm TF}$ is the Thomas-Fermi screening parameter, and $\varepsilon_\infty$ is the static dielectric screening parameter. Based on the screened Coulomb potential $v_{\rm sc}$, interaction matrix elements can be evaluated by
\begin{equation}
\label{eq:int}
U_{ijkl}=\int\int{\varphi_i^\ast(\rr)\varphi_j^\ast(\rr')v_{\rm sc}(\rr,\rr')\varphi_k{(\rr)\varphi}_l(\rr')\dd\rr \dd\rr'},
\end{equation}
where $\varphi_{i}$ is the local orbital describing the localization of the electron. 

Hybrid functionals combine the local density approximation (LDA) or generalized gradient approximation (GGA) exchange and correlation functional with a fraction of non-local Fock exchange:
\begin{equation}
\label{eq:hybrid}
E^{\rm Hyb}_{\rm xc} = E^{\rm LDA/GGA}_{\rm c} + \alpha E^{\rm Fock}_{\rm x} + (1-\alpha) E^{\rm LDA/GGA}_{\rm x},
\end{equation}
where correlation is retained at the level of LDA/GGA $E^{\rm LDA/GGA}_{\rm c}$ and exchange is balanced between exact Fock $E^{\rm Fock}_{\rm x}$ and standard XC functionals $E^{\rm LDA/GGA}_{\rm x}$ through $\alpha$. The crucial mixing parameter $\alpha$ generally be determined either by theoretical analysis based on the adiabatic connection formalism \cite{Perdew1996}, or by fitting to experimental data \cite{Becke1996}. In previous studies, $\alpha$ was typically kept as a fixed value under pressure. In this work, we obtained $\alpha$ also from DSCC calculations, $\alpha=\varepsilon_\infty^{-1}$, from the perspective of many-body perturbation theory\cite{Zhang2020}. In the present work, all calculations are based on the Heyd-Scuseria-Ernzerhof (HSE) hybrid functional \cite{Heyd2003}, taking parameter $\omega$ controls the decomposition of the Coulomb kernel into short-range and long-range contributions to the exchange. Unless otherwise specified, HSE03 refers to the HSE hybrid functional with $\omega = 0.3$ \AA$^{-1}$ and $\alpha=0.25$. We also refer to the HSE hybrid functional with the same $\omega$, and a fixed $\alpha$ taking other values within the whole range of pressures as HSE03. For $\omega = 0.3$ \AA$^{-1}$, and $\alpha=\varepsilon_\infty^{-1}$, we denote it as $\alpha_{\rm sc}$-HSE03 in the following.

All the calculations are performed in the Vienna ${ab~initio}$ simulation package (VASP) \cite{Kresse1996,Kresse1999}, which is based on projector-augmented wave (PAW) framework \cite{Blochl1994,Kresse1999,Fang2019}.  To investigate the influence of magnetic ordering, we considered both FM order, and AFM order along the [111] direction. It should be noted that recent study has proposed a scheme for PM state calculations within the static mean-field framework \cite{Trimarchi2018}. Nevertheless, this approach involves expensive supercell calculations and sophisticated disorder operations.
We only focus on AFM and FM orderings to capture the essential magnetic effects in this work.
Experimental evidence has demonstrated that AFM MnO and NiO exhibit a distorted rhombohedral structure with slight contraction along the trigonal (111) axis of the cubic structure \cite{Yoo2005,Gavriliuk2023}. There is also a structural transformation from the distorted $B1$ (NaCl) phase to the $B8$ (NiAs) structure at around 90 GPa for MnO. In this study, we neglect the distortion induced by antiferromagnetic interactions and the $B1$-$B8$ structural transformation, only considering the rhombohedral supercell of the NaCl-type structure, with two formula units of MnO or NiO.  
For NiO, simulations range from 0 to 1000 GPa, while for MnO, given its experimentally observed insulator-metal transition pressure around 90–105 GPa, simulations were conducted up to 300 GPa. 

The exchange and correlation functional is given by the GGA of Perdew-Burke-Ernzerhof \cite{PBE1996} (PBE) form. Unless otherwise specified, in DFT+$U$ calculations with fixed parameters, we take $U$=8 eV, $J$=1 eV for NiO, and $U$=5.5 eV, $J$=1 eV for MnO. 
In DFT+DSCC scheme, the on-site Coulomb interaction parameters are pressure-dependent and determined by interaction matrix elements evaluated by DSCC. In $\alpha_{\rm sc}$-HSE03 scheme, the mixing parameters are also pressure dependent, taken as $\varepsilon^{-1}_{\infty}$ evaluated by DSCC.
The cut-off energy of 500 eV for the plane-wave basis. The pseudopotential has a critical impact in high-pressure calculations \cite{Zhang2021}. we used PAW pseudopotentials Mn\_sv, Ni\_pv, and standard O. Specifically, valence electrons included 2$s$ and 2$p$ for O, 3$s$, 3$p$, 3$d$ and 4$s$ for Mn, and 3$p$, 3$d$ and 4$s$ for Ni. Monkhorst-Pack\cite{Monkhorst1976} $k$-point meshes were set as $6\times 6 \times 6$. Structural relaxations under fixed pressure employed convergence criteria of atomic forces below 0.01 eV/\r{A}, with SCF convergence thresholds of $1\times10^{-6}$ eV for total energy differences. All the DFT+$U$ calculations were performed using anisotropic Liechtenstein formalism.
The fully localized limit form \cite{Anisimov1993} was employed in order to account for the double counting term in the Hubbard correction.

\

\section{Data Availability}

The data that support the findings of this study are available from the corresponding
author upon reasonable request.

\section{Acknowledgement}
We thank Yu Liu for helpful discussions. This work was supported by the National Key Research and Development Program of China (Grant No. 2021YFB3501503), the National Natural Science Foundation of China (Grants No. U23A20537 and No. U2230401), and Funding of National Key Laboratory of Computational Physics. We thank the Tianhe platforms at the National Supercomputer Center in Tianjin.

\small

\bibliographystyle{unsrt}

\bibliography{bib}

\end{document}